\documentclass[aip, apl, onecolumn]{revtex4-1}
\usepackage{bm}
\draft 
\usepackage{graphicx}
\usepackage{dcolumn}
\usepackage{amsmath}
\usepackage[utf8]{inputenc}
\usepackage[T1]{fontenc}
\usepackage{mathptmx}
\usepackage{etoolbox}
\usepackage{color}
\usepackage{ulem}
\begin{document}

\title{\textcolor{blue}{Alternative harmonic detection approach for quantitative determination of spin and orbital torques}} 

\author{Y. Xu}
\author{B. Bony}
\author{S. Krishnia}
 \altaffiliation[Present address: ]{Institute of Physics, Johannes Gutenberg-University Mainz, Mainz, 55128, Germany}
\author{R. Torrão Victor}
\author{S. Collin}
\author{A. Fert}
\author{J.-M. George}
\author{\\V. Cros}
\author{H. Jaffrès}
 \email{henri.jaffres@cnrs-thales.fr}
\affiliation{Laboratoire Albert Fert, CNRS, Thales, Université Paris-Saclay, 91767 Palaiseau, France}


\date{\today}

\begin{abstract}
In this study, the spin-orbit torque (SOT) in light metal oxide systems is investigated using an experimental approach based on harmonic Hall voltage techniques in out-of-plane (OOP) angular geometry for samples with in-plane magnetic anisotropy. In parallel, an analytical derivation of this alternative OOP harmonic Hall detection geometry has been developed, followed by experimental validation to extract SOT effective fields. In addition to accurately quantifying SOT, this method allows complete characterization of thermoelectric effects, opening promising avenues for accurate SOT characterization in related systems. In particular, this study corroborates the critical role of naturally oxidized copper interfaced with metallic Cu in the generation of orbital current in Co(2)|Pt(4)|CuO\(_{\text{x}}\)(3), demonstrating a two-fold increase in damping-like torques compared to a reference sample with an oxidized Al capping layer. These findings offer promising directions for future research on the application aspect of non-equilibrium orbital angular momentum.
\end{abstract}

\maketitle 

In recent years, the efficient exploitation of the electron spin has been central to spintronics for advanced electrical control and manipulation of magnetization. In these perspectives, spin-orbit torques (SOT)\cite{miron2011perpendicular,Liu2012,liu2012science,RevModPhys.91.035004} allowed by conversion of electrical charges into spins inducing magnetic dynamics through angular momentum transfer, have been recognized as one of the most promising approaches to realize the next generation of magnetic random access memory (SOT-MRAM)\cite{acsMRAM,mramAPL}. However, it requires a deep understanding of the underlying physics of SOTs\cite{kim2013layer,PhysRevLett.128.217702NiO,grimaldi2020single,doi:10.1021/acs.nanolett.2c05091-Sachin} and a significantly increased charge-to-spin conversion efficiency\cite{fan2014magnetization,hibino2021giant,lee2021efficient,PhysRevLett.126.107204ZHU}. In heavy metal (HM) / ferromagnet (FM) heterostructures, charge-to-spin conversion can arise from both bulk and interfacial spin-orbit interactions (SOIs), namely the spin Hall effect\cite{sinova2015spin,Liu2012,liu2012science} (SHE) and the Rashba-Edelstein effect (REE)\cite{bychkov1984properties,edelstein1990spin,mihai2010current,manchon2015new}. It has been widely accepted that these effects can lead to damping-like torques (DLT) and field-like torques (FLT)\cite{RevModPhys.91.035004,garello2013symmetry}. On the one hand, SHE is believed to be dominant in the DLT\cite{Liu2012,PhysRevB.66.014407anatomy,PhysRevB.89.214419FLDLMgO} expressed as $\bm{\tau}_{\text{DL}} \sim \hat{\bm{m}} \times (\hat{\bm{m}} \times \hat{\bm{\sigma}})$, where $\hat{\bm{\sigma}}$ is the spin polarization vector and $\hat{\bm{m}} \times \hat{\bm{\sigma}}$ is parallel to the effective damping-like field ($H_{\text{DL}}$).
On the other hand, interfacial REE originating from symmetry-breaking at interfaces and subsequent spin-splitting in the electronic band structure plays a crucial role in generating FLT\cite{kim2013layer,miron2011perpendicular}, namely $\bm{\tau}_{\text{FL}} \sim \hat{\bm{m}} \times \hat{\bm{\sigma}}$, in which the effective field-like field ($H_{\text{FL}}$) component is parallel to $\hat{\bm{\sigma}}$. 
To overcome the limitations of industrial applications, new materials and underlying physical mechanisms are urgently required to lower the critical current for switching.

Recently, orbital angular momentum (OAM), the previously underestimated degree of freedom, has been found to respond to charge current stimuli in a weak spin-orbit coupling (SOC) system\cite{PhysRevB.103.L020407OAM,choi2023observation,go2018intrinsic,Lee2024,el2023observation,go2021orbital}. Two counterpart phenomena, the so-called orbital Hall effect (OHE)\cite{choi2023observation,PhysRevLett.131.156702MO-OHE,APL_OHE} and the orbital Rashba-Edelstein effect (OREE)\cite{PhysRevLett.128.067201OREM,PhysRevLett.125.177201orbitalSOT,PhysRevB.103.L020407OAM} have been proposed to efficiently realize orbital torque \textit{via} OAM currents. Those may be generated in light metals through OHE\cite{choi2023observation,PhysRevLett.131.156702MO-OHE} diffusing then into an adjacent HM layer or FM layer where it is converted into spin by the SOI, and then applying a torque on the FM magnetization by analogy to the SOT\cite{Lee2021_orbitalSOT,PhysRevB.109.014420_INTF_CuO}. Another mechanism for orbital torques is the OREE correlated to specific interfaces of light metals and their oxidized form\cite{NP_IOREE,PhysRevLett.125.177201orbitalSOT,PhysRevB.109.014420_OREE_INTf}, especially in contact with an adjacent FM layer\cite{doi:10.1021/acs.nanolett.2c05091-Sachin}. Recent works have shown that SOT efficiency can be greatly increased by harnessing OHE, making further progress toward the realization of SOT-MRAM\cite{Gambardella_FIMOT,PhysRevMaterials.7.L111401}. In particular, in our previous work on Co(2)/Pt(t$_{Pt}$)/CuOx(3) multilayers with in-plane magnetic anisotropy (IMA)\cite{10.1063/5.0198970_APLM}, it exhibits a significant DLT up to 2.38~mT/(10\(^{11}\) A/m\(^2\)).The most pronounced effect is observed with a t$_{Pt}\simeq$~4~nm thick Pt insertion layer. Our work emphasized the critical role of OAM for the benefit of a more efficient charge-spin conversion.

Here, we have developed and experimentally validated angular harmonic Hall measurements in the out-of-plane (OOP) geometry for two in-plane magnetized samples consisting of two different oxidized light metals, either CuO\(_{\text{x}}\) or AlO\(_{\text{x}}\), different in that the former provides additional orbital torque.
The interest in the angular OOP measurement approach prevails because it offers distinct advantages over the commonly used in-plane (IP) geometry\cite{thermalIP}. First, we demonstrate that, unlike the IP geometry, a quantitative determination of the torques does not require a complete set of measurements at different magnetic fields. This is due to the distinct angular signature from the thermal effects. Second, the proposed OOP method provides greater accuracy in properly separating and quantifying the field-like and damping-like torque components if the planar Hall (PHE) and anomalous Hall (AHE) effects are of the same order of magnitude due to their distinct responses. This is particularly true when one considers electron-magnon interactions that possibly provide spurious magnon magnetoresistance (MMR) terms admitting the same $\cos^3\left(\phi\right)$ signature as the FLT in the azimuthal configuration~\cite{Noel1,Noel2} and thus leading to necessary corrections to the $V_{2\omega}$ signal. In the OOP, MMR should be discarded, the angular momentum (spin, orbital momentum) injected into the ferromagnet being always transverse to the magnetization. The counterpart of angular OOP is the possible occurrence of additional in-plane thermoelectric effects \cite{PhysRevB.102.024427_thermalSOT} with respect to an in-plane temperature gradient, which can also be accurately examined and disentangled. To verify its applicability, we first measure identical samples providing both spin and orbital torque response as the one studied in Krishnia \textit{et. al}~[\onlinecite{10.1063/5.0198970_APLM}], which is SiO\(_2\)(sub.)|Co(2)|Pt(4)|CuO\(_{\text{x}}\)(3) (number in parentheses is the thickness in units of nanometers). Moreover, a reference sample of Co(2)|Pt(4)|AlO\(_{\text{x}}\)(3) free of any orbital contribution, only integrating Co|Pt SHE source, is used for direct comparison~\cite{10.1063/5.0198970_APLM}. 

\begin{figure}
    \centering
    \includegraphics[width=0.7\linewidth]{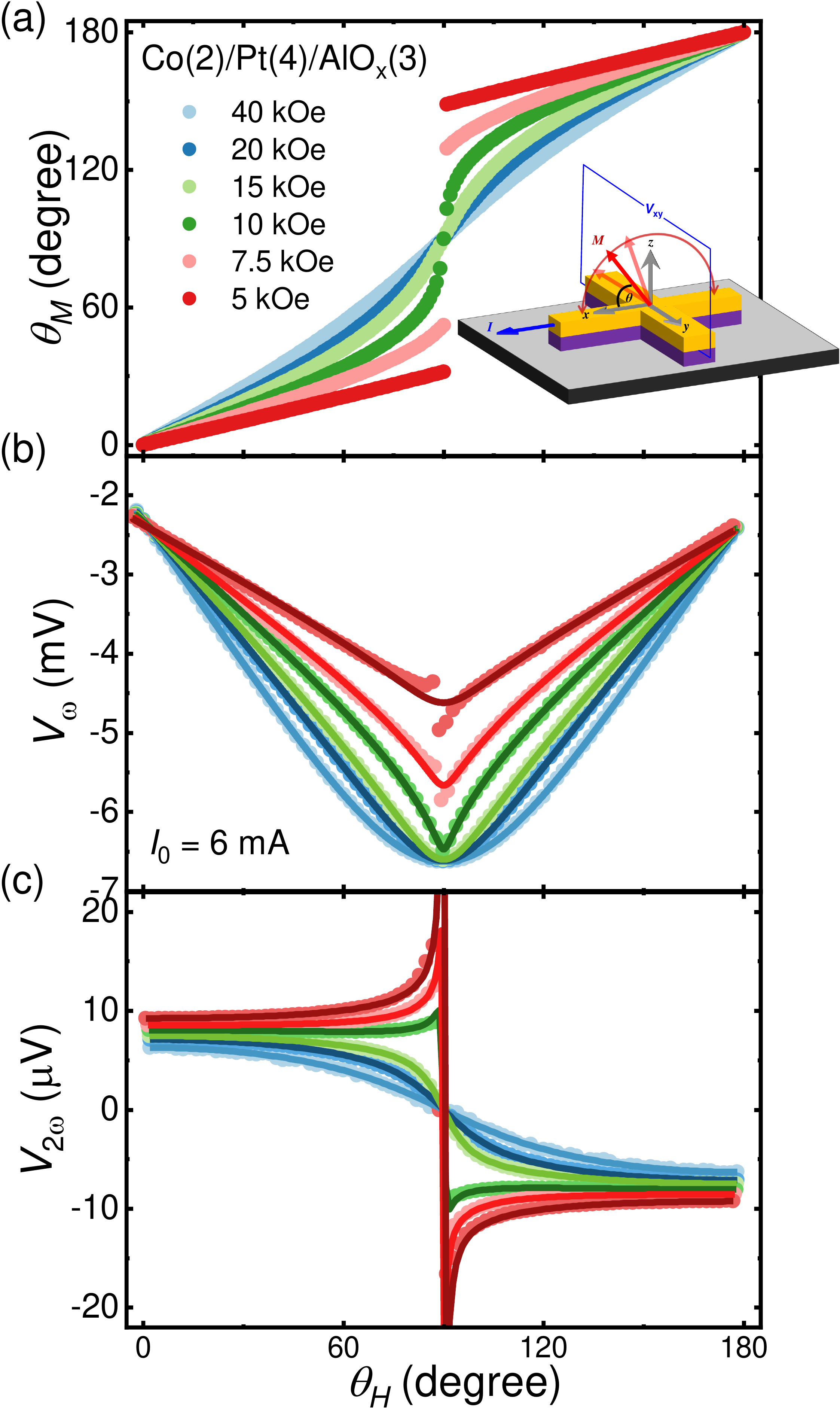}
    \caption{(a) $\theta_M$ \textit{vs.} $\theta_H$ in the OOP geometry which depicts the stable orientation of $\mathbf{M}$ with magnetization switching at $\theta_H=\frac{\pi}{2}$ for $\frac{H_K}{2}<H_{ext}<H_K$. The inset shows the schematics of the device designed for the electrical and torque measurements and the geometry conventions. (b) and (c) Angular-dependent 1$^\text{st}$ and 2$^\text{nd}$ harmonic Hall voltages (color dots) measured with an AC of 6~mA under different $H_{\text{ext}}$ and corresponding fitting curves (colored lines). Experiments are performed at 300~K.}
    \label{fig:1}
\end{figure}

These two film stacks are grown at room temperature using DC magnetron sputtering onto thermally oxidized Si|SiO\(_2\) substrates. The capping layers of Cu and Al are naturally oxidized in the air after deposition, thus forming, respectively, Cu | CuO\(_{\text{x}}\) and AlO\(_{\text{x}}\) interfaces. 
\begin{table}
\caption{\label{tab:R} AHE, PHE and effective anisotropy field for each sample }
\begin{ruledtabular}
\begin{tabular}{lccr}
Sample&$R_{AHE}$($\Omega$)&$R_{PHE}$($\Omega$)&\( H_k \)(kOe)\\
\hline
Co(2)/Pt(4)/AlO\(_{\text{x}}\)(3) & 0.72 & 0.138 & 9.5\\
Co(2)/Pt(4)/CuO\(_{\text{x}}\)(3) & 0.4 & 0.08 & 9.6\\
\end{tabular}
\end{ruledtabular}
\end{table}
Anomalous Hall (AHE) and planar Hall effects (PHE) for these two samples, essential for the signal-to-noise ratio of the harmonic signals, are gathered in Table~I as well as their effective in-plane anisotropy field $H_K$. For the exact same Co(2)|Pt(4) active layer part, differences in the corresponding AHE and PHE values indicate the current shunt in CuOx(3) due to the presence of an effective 1~nm thick remaining Cu metallic character at the interface with Pt. To perform electrical and torque measurements, 5~\textmu m Hall bars have been fabricated from the stacks using ultra-violet lithography and Ar\(^{\text{+}}\) ion milling technique (inset of Fig.~\ref{fig:1}(a)). To properly estimate the shunting current in the Pt layer, a simple parallel resistor model of the film stack is included. We measured the longitudinal resistivity of the Hall bars in various thicknesses of Pt, the dependent relation of which can produce the resistivity of Pt and the rest of the layers in the stack (see the supplementary information of Ref.~[\onlinecite{10.1063/5.0198970_APLM}]). Using this method, the resistivity of Pt is estimated to be 25~$\mu\Omega\cdot \text{cm}$. Thus, the shunting current in the Pt layer of any stack can be calculated by measuring the Hall bars longitudinal resistance. The readers may refer to Ref.~[\onlinecite{10.1063/5.0198970_APLM}] for more details.

We first model the 2$^\text{nd}$ harmonic OOP response for IMA samples, acquired while rotating the magnetization in the \(xz\)-plane (Fig.~\ref{fig:1}(a)). The magnetic energy of the system is
\begin{equation}
-\epsilon = -\left(2 \pi M_S^2 - \frac{K_S}{t}\right) \cdot \left(\frac{M_z}{M_S}\right)^2 + H_x M_x + H_z M_z,
\label{eq:1a}
\end{equation}
where \(K_{\text{S}}\) is the surface energy density of the perpendicular magnetic anisotropy (PMA) introduced by the SOI, and \(t\) is the FM layer thickness. \(M_{\text{x, z}}\) (\(H_{\text{x, z}}\)) are the IP and OOP components of magnetization, $M_S$ (external magnetic field, \(H_{ext}\)) respectively. The first term in parenthesis in eq.~(\ref{eq:1a}) represents the effective in-plane anisotropy energy. It can be put in the form \((H_k \cdot M_S) / 2\), where \(H_k\) represents the effective in-plane anisotropy field. The equilibrium magnetization direction corresponds minimizing the energy Eq.~(\ref{eq:1a}) with respect to \(M_{\text{z}}\), thus giving the equilibrium solution which reads $\frac{M_z}{M_S} = \frac{H_z}{H_{\text{eff}}}$
with $H_z=H_{ext} \sin\left(\theta_H\right)$, and $H_{\text{eff}}=H_{\text{ext}} \cdot \cos(\theta_H - \theta_M) \cdot \frac{d\theta_H}{d\theta_M}=H_{ext} \cos(\theta_H - \theta_M)+H_K\cos\left(2\theta_M\right)$ is the effective field magnitude acting on the magnetization vector, which stabilizes it in its equilibrium state ($\theta_M$). The minimization procedure can also be written in terms of the Euler angles, respectively \( (\theta_M, \phi_M) \) and \( (\theta_H, \phi_H) \), to yield:
\begin{equation}
\tan{\theta_M} = \frac{H_{\text{ext}} \cdot \sin{\theta_H}}{H_k \cdot \cos{\theta_M} + H_{\text{ext}} \cdot \cos{\theta_H}}
\label{eq:2}
\end{equation}
Eq.~(\ref{eq:2}) can be solved self-consistently to obtain the implicit function of \(\theta_M\)(\(\theta_H\)) and the effective anisotropy field \(H_{\text{eff}}\). 

For the Hall bar in the inset of Fig.~\ref{fig:1}(a) with current injected in the \(\hat{\bm{x}}\) direction, the SOT effective fields are given by: the DL term \( \bm{H}_{\text{DL}} = H_{\text{DL}} \hat{\bm{m}} \times \hat{\bm{y}} \) and the FL term \( \bm{H}_{\text{FL}} = H_{\text{FL}} \hat{\bm{y}} \), where \(H_\text{DL}\) and \(H_\text{FL}\) are the effective field strength proportional to the current \( I = I_0 \sin(\omega t) \), and \(\hat{\bm{y}}\) is the unit vector parallel to the spin vector. In the OOP geometry, $H_{\text{DL}}$ and $H_{\text{FL}}$ write \( \bm{H}_{\text{DL}} = (H_{\text{DL}} \sin(\omega t)) \hat{\bm{\theta}} \) and \( \bm{H}_{\text{FL}} = (H_{\text{FL}} \sin(\omega t)) \hat{\bm{\phi}} \), where $\hat{\bm{\theta}}$ and $\hat{\bm{\phi}}$ are the polar and azimuthal unit vectors, respectively. The effective SOT fields synchronize with the low-frequency alternating current \( I \), causing the magnetization to oscillate around its equilibrium position \((\theta_M, \phi_M)\) with time-dependent variations \((\Delta \theta_M, \Delta \phi_M)\)\cite{harmonicSOT}. This small modulation is proportional to the oscillating SOT fields so that it can be calculated by its derivative given by
$\Delta \theta_M = \frac{d\theta_M}{d \bm{H}_I} \Delta \bm{H}_I = \frac{d\theta_M}{d \bm{H}_I^\theta} \Delta \bm{H}_I^\theta$
and
$\Delta \phi_M = \frac{d\phi_M}{d \bm{H}_I} \Delta \bm{H}_I = \frac{d\phi_0}{d \bm{H}_I^\phi} \Delta \bm{H}_I^\phi$ 
where \( \Delta \bm{H}_I = \bm{H}_{\text{DL}} + \bm{H}_{\text{FL}} + \bm{H}_{\text{Oe}} \) is the total effective field induced by the current, in which \( \bm{H}_{\text{Oe}} \) is the Oersted field due to current flowing in the HM layer. Because the variations of \( \theta_M \) and \( \phi_M \) depend only on the polar and azimuthal components of \( \Delta H_I \) respectively, the derivatives can be replaced with \( H_I^\theta \) and \( H_I^\phi \). To calculate these derivatives, one can exploit the principle that the magnetic field dependence of magnetization orientation is irrelevant to the nature of the field~\cite{PhysRevB.89.214419FLDLMgO}, to obtain:
\begin{equation}
\Delta \theta_M = \frac{H_{\text{DL}} \sin(\omega t)}{H_{\text{ext}} \cdot \cos(\theta_H - \theta_M)} \cdot \frac{d\theta_M}{d\theta_H}, \label{eq:delta2}
\end{equation}
and
\begin{equation}
\Delta \phi_M = \frac{(H_{\text{FL}} + H_{\text{Oe}}) \sin(\omega t)}{H_{\text{ext}} \cdot \cos(\theta_H)} \label{eq:phi_0}.
\end{equation}
\begin{figure}
    \centering
    \includegraphics[width=1\linewidth]{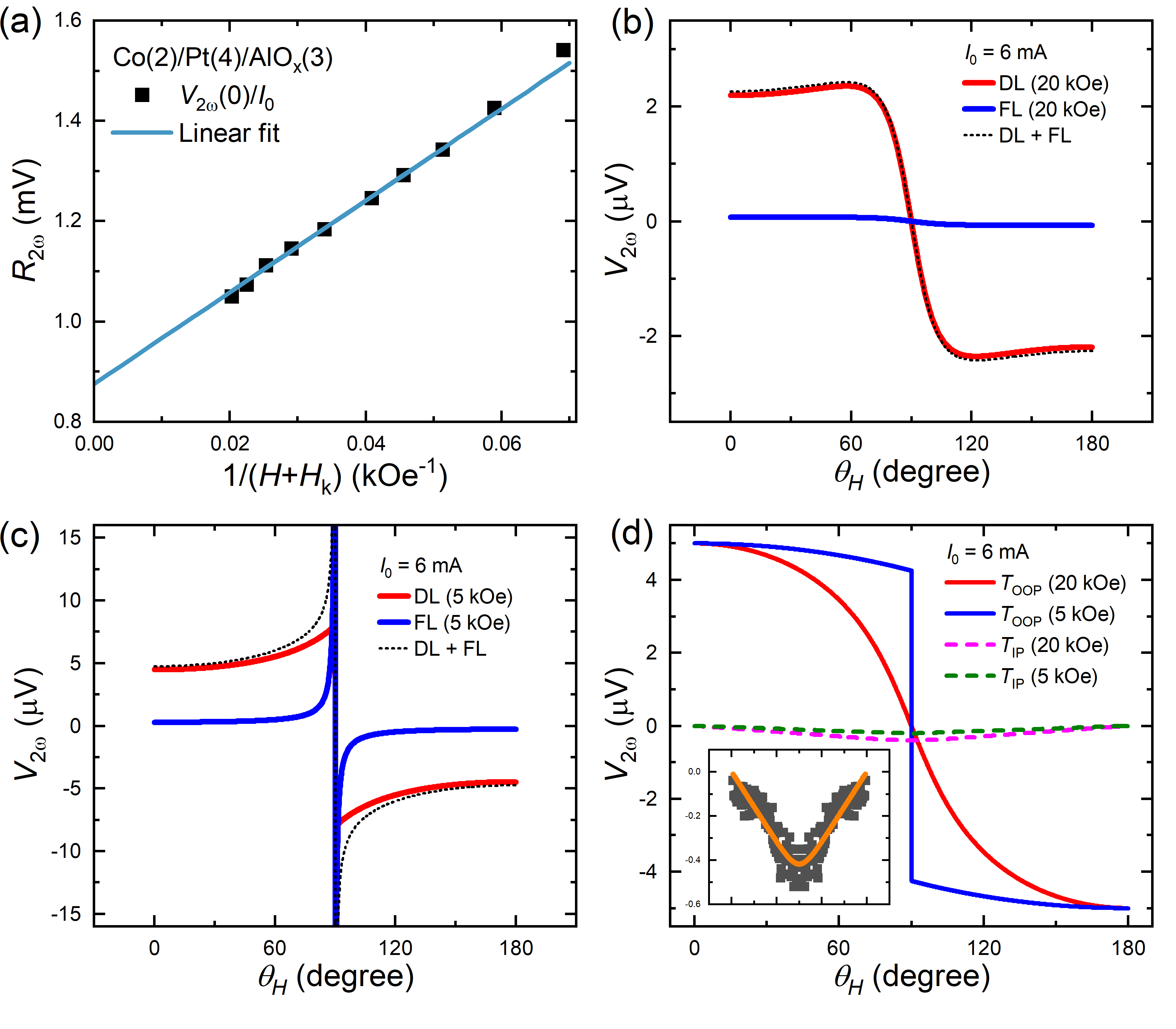}
    \caption{(a) 2$^\text{nd}$ harmonic Hall resistance for Co(2)/Pt(4)/AlO\(_{\text{x}}\)(3), defined as \( R_{2\omega} = V_{2\omega}/I_0 \) with I$_0=6$~mA, plotted \textit{vs.} \( 1/(H_{\text{ext}} + H_k) \) with a linear fit. (b) and (c) display the DL (red) and FL (blue) components of the torque signals decomposed from the 2$^\text{nd}$ harmonic signals measured under external fields of 20~kOe and 5~kOe with peak current of 6 mA. The black dotted lines represent the total torque signals comprising both components. (d) The remaining thermoelectric components of the \(T_{\text{OOP}}\) and (\(T_{\text{IP}}\)) terms, as shown by the colored solid and dashed lines.  The inset in (c) presents the symmetrical part of the raw data and the fitting curve of the $T_{\text{IP}}$ signal measured at 20~kOe. Experiments were performed at 300~K.}
    \label{fig:2}
\end{figure}

Eq.~(\ref{eq:phi_0}) follows a similar rationale and is valid in the IMA geometry. These modulations can be observed as measurable changes in the harmonic Hall voltage signals. The total Hall resistance consists of two components: the anomalous Hall effect (AHE) and the planar Hall effect (PHE), expressed as: 
\begin{equation}
V_{\text{Hall}} = I_0 \sin(\omega t) \cdot \left( R_{\text{AHE}} \sin(\theta_M) + R_{\text{PHE}} \cos^2(\theta_0) \sin(2\phi_M) \right),
\label{eq:Hall}
\end{equation}
where \( R_{\text{AHE}} \) and \( R_{\text{PHE}} \) are the amplitudes of the AHE and PHE resistances. We have then
\begin{align}
V_{\text{Hall}} &= I_0 \sin(\omega t) \cdot \left[ R_{\text{AHE}} \sin\left( \theta_M + \Delta \theta_M \right) \right. \nonumber \\
& \quad + R_{\text{PHE}} \cos^2\left( \theta_M + \Delta \theta_M \right) \sin\left( 2(\phi_M + \Delta \phi_M) \right) \left. \right] \label{eq:V_Hall}
\end{align}
Expanding Eq.~(\ref{eq:Hall}) to the zero and first order of the Taylor series at \( (\theta_M, \phi_M) \) and utilizing the results from Eqs.~(\ref{eq:delta2})-(\ref{eq:V_Hall}) yields the expressions for the 1$^{\text{st}}$ and 2$^\text{nd}$ harmonic Hall signals, with \( \phi_M \) set to zero due to the experimental geometry:
\begin{equation}
V_\omega = I_0 R_{\text{AHE}} \sin(\theta_M) \label{eq:V_omega},
\end{equation}
and
\begin{align}
V_{2\omega} &= -\frac{1}{2} \frac{dV_\omega}{d\theta_M} \frac{H_{DL}}{\left[H_{ext} \cos(\theta_H - \theta_M)+H_K\cos\left(2\theta_M\right)\right]} \nonumber \\
& \quad - I_0 R_{\text{PHE}} \cos^2(\theta_M) \frac{ H_{FL} + H_{OE}}{H_{ext} \cos(\theta_H)} \nonumber \\
& \quad + I_0 \alpha \left( T_{\text{OOP}} \cos(\theta_M) + T_{\text{IP}} \sin(\theta_M) \right) \label{eq:V_2omega_2}
\end{align}
Eq.~(\ref{eq:delta2}) requires the information from the 1$^\text{st}$ harmonic signal and later knowledge of $H_{\text{eff}}$. The last term in Eq.~(\ref{eq:V_2omega_2}) is the thermoelectric contribution considering the anomalous Nernst effect (ANE)\cite{thermalIP}, in which $T_{\text{OOP}}$ and $T_{\text{IP}}$ represent, respectively, the normal and in-plane components of the temperature gradient, with $\alpha$ the ANE coefficient.

We now turn to the experiments with the Co(2)|Pt(4)|AlOx\(_{\text{x}}\)(3) sample free of any orbital effects and subsequent fittings and analyzes. 1$^\text{st}$ and 2$^\text{nd}$ harmonic signals \textit{vs.} \( \theta_H \) are measured simultaneously using an AC of 19~Hz and peak amplitude \(I_0\) of 6~mA. We have checked that the $V_{2\omega}$ signals correctly scale as $I_0^2$, as expected. The results are displayed in Figs.~\ref{fig:1}(b) and (c) for different $H_{\text{ext}}$ as indicated by the colored data points. To fit $V_{1\omega}$ (Fig.~\ref{fig:1}(b)) using Eq.~(\ref{eq:V_omega}), the self-consistent Eq.~(\ref{eq:2}) has been numerically solved to establish the \(\theta_M \) \textit{vs.} \( \theta_H \) relationship. The latter is reported on Fig.~\ref{fig:1}(a) for different values of  $H_{\text{ext}}$ values and for H$_k\simeq$ 9.47~kOe, consistent with our previous measurements (10~kOe) for similar Co(2)|Pt(4)|AlO\(_{\text{x}}\)(3) samples~(Ref.[\onlinecite{10.1063/5.0198970_APLM}]). This value of \( H_k \) was subsequently used to fit the 2$^\text{nd}$ harmonic signals (Fig.~\ref{fig:1}(c)). In particular, only the symmetric part of \( V_{2\omega} \) around the 90$^o$ polar angle, although small, is assigned to the term \( T_{\text{IP}} \). The remaining antisymmetric components, shown in Fig.~\ref{fig:1}(c), are then fitted with three free parameters: $H_{\text{DL}}$, $H_{\text{FL}}$, and \( T_{\text{OOP}} \). Moreover, one observes a specific signature of V$_{2\omega}$ in the "low" $H_{\text{ext}}$ region, which is characterized by an increase of |V$_{2\omega}$| \textit{vs.} $\theta_H$ followed by an abrupt drop (or either jump) from positive to negative values. This drop corresponds to the magnetization reversal driven by $H_{\text{ext}}$ toward the other branch for greater energy stability.
From the energy minimization procedure, one can show that the observed increase in V$_{2\omega}$ close to the origin ($\theta_H,\theta_M \rightarrow 0$) has to be correlated with a positive curvature of the DLT response sensitivity $S\propto R_{AHE}\left(\frac{H_{DL}\cos\left(\theta_M\right)}{H_{eff}\left(\theta_M\right)}\right)$ \textit{vs.} $\theta_H$ in the field range $H_{ext}<\left(\frac{3+\sqrt{13}}{2}\right) H_K\simeq 33$~kOe. Such increase of V$_{2\omega}$ is exemplified in the data reported in Fig.~\ref{fig:2}(b)-(c). This witness the direct action of the DLT on V$_{2\omega}$ which cannot be explained by any of the two thermal effects (neither $T_{IP}$ nor $T_{OOP}$) giving an opposite curvature response. Moreover, we expect a magnetization reversal (or switching) in the vicinity of $\theta_H\simeq \frac{\pi}{2}$ during the rotation process for $\frac{H_K}{2}<H_{ext}<H_K$ (at 0~K). This is reproduced in our simulations (Fig.~\ref{fig:1}(a)) and in the experimental data (Fig.~\ref{fig:1}(c)). Such irreversible switching for the OOP normally occurs for $\sin\left(2\theta_M\right)=\sqrt{\frac{4}{3}\left(1-\frac{H_{ext}^2}{H_K^2}\right)}$ or $\sin\left(\theta_H-\theta_M\right)=\sqrt{\frac{1}{3}\left(\frac{H_{K}^2}{H_{ext}^2}-1\right)}$ but can be anticipated around $\theta_H\simeq \frac{\pi}{2}$ considering possibly nucleation of domains by thermal activation.

We now focus on the $V_{2\omega}$ acquired at various $H_{\text{ext}}$ in the sample plane ($\theta_H=0$ and $\theta_M=0$). From Eq.~(\ref{eq:V_2omega_2}), it is obvious that the term \( H_{\text{DL}} \) scales with \( 1/(H_{\text{ext}} + H_k) \), the term \( H_{\text{FL}} \) scales as \( 1/H_{\text{ext}} \), while the thermal contribution remains constant. In Fig.~\ref{fig:2}(a), the extracted amplitude of \( R_{2\omega} = V_{2\omega}/I_0 \) is plotted by $\textit{vs.}$ \(1/(H_{\text{ext}} + H_k)\), showing a strong linear relationship. The linear fit yields an intercept of 5.23~\text{$\mu V$}, corresponding to the thermoelectric voltage contributed by \( T_{\text{OOP}} \) component, while the slope reflects the value of \( H_{\text{DL}} \), approximately 1.18~mT / $\left(10^{11} \text{A/m²}\right)$.
It is reasonable to consider only $H_{\text{DL}}$ and the thermoelectric contributions for fields larger than \( H_k \), while $H_{\text{FL}}$ can be extracted by fitting the signals measured in low fields. Given that the anomalous Hall resistance is roughly 5 times greater than the planar Hall resistance for the Co(2)|Pt(4)|AlO\(_{\text{x}}\)(3) sample (see details in Table \ref{tab:R}), the PHE contribution can be neglected in the high-field regime. 

Based on this analysis, the torque and thermoelectric components can be decomposed from the 2$^\text{nd}$ harmonic signals for different external fields. In Figs.~\ref{fig:2}(b) and (C) the torque signals for external fields of 20 and 5~kOe are further decomposed into contributions from DL (red curve) and FL (blue curve), respectively. The black dotted curves represent the total torque signal, composed of the two components. For $H_{\text{ext}}$ = 20~kOe, the $H_{\text{FL}}$ component is negligibly small because of the small magnitude of $R_{\text{PHE}}$, and the $H_{\text{DL}}$ component aligns with the full torque curve. In contrast, at 5~kOe, the full torque curve differs noticeably from the $H_{\text{DL}}$ curve, indicating a considerable contribution from $H_{\text{FL}}$ with distinct function curvature, allowing the extraction of \(H_{\text{FL}} \). As shown in Fig.~\ref{fig:2}(d), the \(T_{\text{OOP}} \) signal, which is antisymmetric, is one order of magnitude larger than that of \(T_{\text{IP}} \). The inset of Fig.~\ref{fig:2}(d) shows the raw data and the fitted curve for the in-plane component of \(T_{\text{IP}} \) for $H_{ext}=$ 20~kOe, which is obtained by extracting the symmetric part of the raw 2$^\text{nd}$ harmonic signals. In Hall bar devices fabricated from metallic ferromagnetic stacks, the large difference between \(T_{\text{OOP}} \) and \(T_{\text{IP}} \) is mainly due to inhomogeneous thermal conductivity. The former is driven by the strong contrast in thermal conductivity between the Si substrate and the surrounding air at the sample surface. In contrast, because of the Hall bar geometrical asymmetry, the in-plane conductivity is more uniformly distributed, with, nonetheless, a finite-temperature gradient introduced by asymmetric heat dissipation between the contacts.

\begin{figure}[hbt!]
    \centering
    \includegraphics[width=0.9\linewidth]{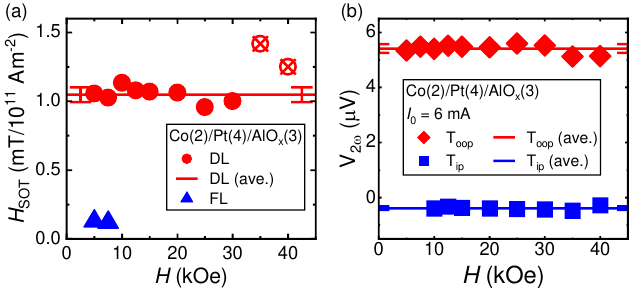}
    \caption{(a) and (b) are the values of torque efficiency and thermoelectric voltage (with injected current amplitude of 6~mA) extracted from the fitting for various $H_{\text{ext}}$. The solid lines are the averaged values of the extracted parameters with error bars indicating the variations. }
    \label{fig:3}
\end{figure}

In Figs.~\ref{fig:3}(a) and (b), we display the different values of the torques and thermal coefficients for all $H_{\text{ext}}$ and compare them. The magnitudes of $H_{\text{DL}}$ remain almost constant throughout the field window, except for those obtained in the largest fields.
Note that the torque signals are quite small at large fields, reducing the signal-to-noise ratio, and consequently, the measurement accuracy. Furthermore, the resemblance between the \(T_{\text{OOP}} \) curve in Fig.~\ref{fig:2}(c) at high field (especially for 40~kOe and 35~kOe) and the $H_{\text{DL}}$ signal reduces the reliability of the fitting process in that region. This argument is reflected in the significant deviation of $H_{\text{DL}}$ extracted from the measurement for $H_{ext}=$ 40 and 35~kOe. Furthermore, in Fig.~\ref{fig:3}(b), the thermal coefficients remain consistent in the different applied magnetic fields. In summary, the averaged values for $H_{\text{DL}}$ and $H_{\text{FL}}$ are 1.05 and 0.74~mT/$\left(10^{11} \text{A/m}^2\right)$, respectively, and the \(T_{\text{OOP}} \) and \(T_{\text{IP}} \) components are 5.40 and -0.38~\text{\(\mu V\)}. Extracted SOT fields are further scaled with respect to the current density in Pt, which is 10$^{11}$ A/m$^2$.
\begin{figure}[hbt!] 
    \centering
    \includegraphics[width=0.7\linewidth]{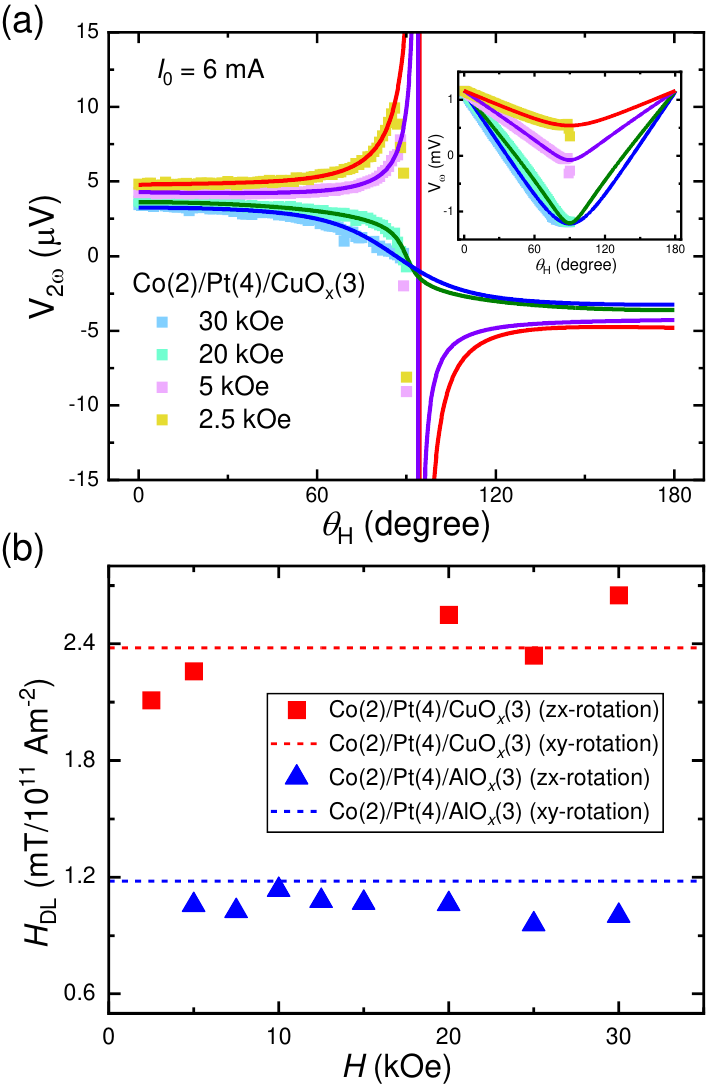}
    \caption{(a) Angular-dependent 2$^{nd}$ (1$^{st}$ in the inset) harmonic raw data (scatters) and fitting curves (colored lines) for the orbital sample of Co(2)|Pt(4)|CuO\(_{\text{x}}\)(3) measured in the OOP geometry with injected current of~6 mA amplitude. (b) Comparison of HDL of the reference and orbital samples obtained by different harmonic measurements in the OOP and IP geometry. }
    \label{fig:4}
\end{figure}

Having confirmed the ability of the OOP angular 2$^\text{nd}$ harmonic measurement method, we have measured and analyzed the characteristic effective SOT generated from the Co(2)|Pt(4)|CuO\(_x\)(3) orbital sample involving an orbital contribution. The results and the fitting are reported in Fig.~\ref{fig:4}(a). The AHE angular variations that vary for different $H_{\text{ext}}$ when $H_{ext}$< \text{or} > $H_K$, are shown in the inset together with the fitting tracking the magnetization angle $\theta_M$.
In Fig.~\ref{fig:4}(b), we compare the values of $H_{\text{DL}}$ taken from the reference (blue) and those resulting from the orbital samples (red) obtained \textit{via} the two different geometries. Dashed lines are the result of Krishnia \textit{ et al.} (Ref.~[\onlinecite{10.1063/5.0198970_APLM}]) of the angular variation geometry of the IP, and the scatter points are the results obtained from OOP showing a very excellent consistency between these two methods. In addition, the twofold increase in $H_{\text{DL}}$ for the orbital sample with an oxidized Cu capping layer, from $H_{\text{DL}}\simeq1.2$~mT/(10$^{11}$~A/m$^2$) to $H_{\text{DL}}\simeq2.4$~mT/(10$^{11}$~A/m$^2$) quantitatively confirms the unique role of the Cu | CuO$_x$ interface in terms of the production of an orbital current. Such SOT enhancement cannot be accounted for from spurious thermal effects, generally well discriminated by the two methods, nor from additional MMR contribution as recently proposed~\cite{Noel1,Noel2} possibly competing with the SOT signal in the IP geometry for a very thin ferromagnetic layer. At last, the thermoelectric voltages measured in Co(2)|Pt(4)|CuO\(_x\)(3) with 6~mA injected current amplitude are estimated to be 2.5 and -0.32~\text{\(\mu V\)}, for the \(T_{\text{OOP}} \) and \(T_{\text{IP}} \) components respectively.

\vspace{0.1in}

In conclusion, we have provided the experimental proof of an alternative harmonic technique consistent with the IP geometry in terms of the parameter extraction of SOT effective fields. Notably, using this geometrical method, we emphasized the critical role of the naturally oxidized copper layer in generating orbital current and orbital torque with a twofold larger DL field than the reference sample. More generally, this method paves the way for further investigations of theoretical aspects, accurate measurements and applications of the magnetic torques resulting from non-equilibrium orbital angular momentum, in particular for materials characterized by large thermoelectric effects as Bi-based compounds and related topological insulators.


\begin{acknowledgments}
This work has been supported by the French National Research Agency under the Project “ORION” ANR-20- CE30-0022-02, ANR grant STORM (ANR-22-CE42-0013-02), by France 2030 government investment plan managed by the French National Research Agency under grant reference PEPR SPIN –ADAGE ANR-22-EXSP-0006, and the European Horizon Europe Framework Programme for Research and Innovation (2021-2027) EC Grant Agreement No. 101129641 (OBELIX). 
\end{acknowledgments}

\bibliography{APL_biblio}

\end{document}